\newcommand{\CLASSINPUTinnersidemargin}{19mm}
\newcommand{\CLASSINPUTtoptextmargin}{25mm}

\documentclass[9pt,conference]{IEEEtran}
\usepackage[table]{xcolor}
\usepackage{cite}
\usepackage{amsmath,amssymb,amsfonts}
\usepackage{graphicx}
\usepackage{makecell}   
\usepackage{diagbox}
\usepackage{multirow}
\usepackage{booktabs}
\usepackage{placeins}
\makeatletter
\let\MYcaption\@makecaption
\makeatother
\usepackage[font=footnotesize,subrefformat=parens]{subcaption}
\makeatletter
\let\@makecaption\MYcaption
\makeatother
\usepackage{tikz}
\usepackage{hhline}
\usepackage{nicefrac, xfrac}
\usepackage{lipsum} 
\usepackage[hidelinks]{hyperref}

\renewcommand{\IEEEtitletopspaceextra}{10mm} 

\makeatletter
\def\@xfootnote[#1]{%
  \protected@xdef\@thefnmark{#1}%
  \@footnotemark\@footnotetext}
\makeatother

\newcommand{\variant}[0]{\hspace{0.1cm}\rotatebox[origin=c]{180}{$\Lsh$}\hspace{0.1cm}}

\newcommand{\nlt}[0]{\\\noalign{\vskip-0.15pt}}  
\newcommand{\margintxt}[2]{#1 \textbf{(}\textit{#2}\textbf{)}}

\newcommand\txtbgbox[2][]{\tikz[overlay]\node[fill=blue!20,inner sep=3pt, anchor=text, rectangle, rounded corners=2mm,#1] {#2};\phantom{#2}}

\newcommand{\derbl}[1]{\txtbgbox[fill=blue!25]{#1}}
\newcommand{\derbg}[1]{\textbf{#1}}

\newcommand{\vth}[1]{\rotatebox[origin=c]{90}{#1}}
\newcommand{\vthml}[1]{\vth{\parbox{2cm}{\centering #1}}}

\newcommand{\tablefontsize}[0]{\normalsize }

\makeatletter
\newcommand\blfootnote[1]{%
  \begingroup
  \renewcommand{\@makefntext}[1]{#1}
  \renewcommand\thefootnote{}\footnote{#1}%
  \addtocounter{footnote}{-1}%
  \endgroup
}
\makeatother

\title{Mamba-based Segmentation Model for Speaker Diarization}

\author{
    \IEEEauthorblockN{Alexis Plaquet\IEEEauthorrefmark{2}, Naohiro Tawara, Marc Delcroix, Shota Horiguchi, Atsushi Ando, and Shoko Araki}
    \IEEEauthorblockA{
        NTT Corporation, Japan
    }
}

\date{June 2024}

\begin{document}

\maketitle

\begin{abstract}
Mamba is a newly proposed architecture which behaves like a recurrent neural network (RNN) with attention-like capabilities. These properties are promising for speaker diarization, as attention-based models have unsuitable memory requirements for long-form audio, and traditional RNN capabilities are too limited.
In this paper, we propose to assess the potential of Mamba for diarization by comparing the state-of-the-art neural segmentation of the pyannote pipeline with our proposed Mamba-based variant. Mamba's stronger processing capabilities allow usage of longer local windows, which significantly improve diarization quality by making the speaker embedding extraction more reliable. We find Mamba to be a superior alternative to both traditional RNN and the tested attention-based model. Our proposed Mamba-based system achieves state-of-the-art performance on three widely used diarization datasets.
\end{abstract}

\begin{IEEEkeywords}
Speaker diarization, end-to-end neural diarization, Mamba, state-space model
\end{IEEEkeywords}

\section{Introduction}

\blfootnote{$^{\dagger}$ IRIT, Université de Toulouse, CNRS, Toulouse INP, UT3, Toulouse, France, JSPS International Research Fellow. This work was done during an internship at NTT.}

Speaker diarization consists of finding ``who spoke when,'' given an audio recording, i.e., the goal is to find the time regions where each speaker is active, with no regard for their exact identities.

Neural networks have become ubiquitous in speech processing, and these last years have seen the rise of end-to-end neural diarization (EEND), which uses neural network-based segmentation models to directly optimize the speaker diarization problem through the permutation-invariant training loss \cite{fujita2019eend}. These approaches directly learn to handle overlapped speech, which is a major challenge for other popular approaches that depend on speaker embedding Vector Clustering (VC) such as VBx~\cite{landini2022vbx}. 

Yet clustering-based diarization has remained competitive as pure EEND approaches only support a fixed maximum number of speakers and gets unreasonably memory-expensive on long-term audio. In particular, hybrid approaches \cite{bredin23pyannote21, kinoshita2021icasp} (also called EEND-VC) use both EEND and clustering methods to combine the strengths of both approaches. They do so by applying an EEND network on smaller fixed-length chunks with a sliding window, followed by a speaker-clustering step to align the identities between all windows (since speaker diarization is speaker-permutation invariant), which are finally aggregated to obtain the diarization output.

EEND models are based on recurrent neural networks (RNN) \cite{fujita2019eend,bredin23pyannote21} and attention-based blocks \cite{fujita2019diarizationselfattention,kinoshita2021icasp}. RNNs are an inherently sequential way to process sequential data, the model sees the data frame by frame and predicts the output while updating an internal memory. The most commonly used type of RNN is the Bidirectional Long Short-Term Memory (BiLSTM). On the other hand, attention-based blocks process the whole input sequence at once and can capture long-term dependencies better than LSTMs which are heavily restricted by their limited memory. However, despite the impressive performance of attention, current attention-based modules disregard the sequentiality of the data as they do not make use of positional embeddings \cite{fujita2019diarizationselfattention,chen2024attention}. We believe that modeling speech as a sequence can be beneficial for diarization if it can be combined with strong long-term dependency modeling.

These last years, a new family of models based on State Space Models (SSM) has emerged, in particular, the recently proposed Mamba \cite{mamba, mamba2} has reached performance on par with attention-based models. Mamba acts as an RNN and processes the data sequentially, but possesses much better processing power and memory. This makes Mamba a promising candidate to replace LSTM and compete with attention-based architectures using sequential processing, which might be key to achieving finer diarization of complex regions. Moreover, the time complexity of Mamba (and all RNNs) only increases linearly with sequence length, unlike attention-based models, which have quadratic time complexity.

In order to evaluate the potential of Mamba for diarization, we study its impact on the EEND-VC approach using the open-source \texttt{pyannote.audio} pipeline, which up to now has relied on BiLSTM for its state-of-the-art results~\cite{bredin23pyannote21, plaquet23powerset, baroudi2023pyannote}. In particular, we propose a novel speaker segmentation model, which replaces BiLSTMs with Mamba blocks. Moreover, we take a more fine-grained look at the impact of the window size and the training criterion (multilabel or multiclass~\cite{plaquet23powerset}) in relation to the LSTM-based architecture and the proposed Mamba-based architectures.

In the spirit of reproducible research, we publicly release the code, model outputs and dataset splits of our experiments at \texttt{\small github.com/nttcslab-sp/mamba-diarization}.

\section{Related work}





Multiple recent studies have shown the potential of Mamba for speech applications and confirmed its effectiveness on audio tasks. However, to our knowledge, this is the first paper to investigate the usage of Mamba for speaker diarization.

Bidirectional Mamba was proposed for vision in \cite{zhu2024visionmamba} and used for speech separation in \cite{li2024spmamba, jiang2024dualpathmamba} where it achieves competitive performance with Transformers using fewer parameters. \cite{zhang2024mambainspeech} studied the effectiveness of two bidirectional Mamba architectures and compared them to Transformers architectures on the speech enhancement and speech recognition tasks. It found Mamba to perform better than Transformer and Conformer architectures on both tasks. 

\cite{erol2024audiomamba,shams2024ssamba,yadav2024audiomambaselectivestate} proposed Mamba-based architectures for learning audio representations. We did preliminary experiments with pre-trained models from~\cite{yadav2024audiomambaselectivestate} as our local EEND feature extractor but did not manage to obtain competitive results.

\section{Proposed system}

\subsection{Diarization pipeline}

We study the impact of multiple factors on an EEND-VC pipeline. The pipeline is composed of a local EEND segmentation model, which is our main focus, followed by an embedding extraction and clustering phase. 

\subsubsection{Local EEND segmentation model architecture}

\begin{figure}[tb]
    \centering
    \begin{subfigure}[b]{0.495\linewidth}
        \centering
        \includegraphics[width=\textwidth]{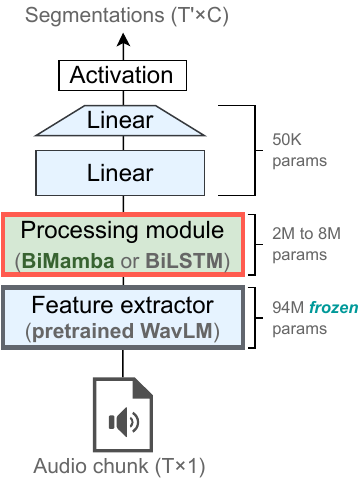}
        \caption{pyannote.audio's local EEND segmentation model architecture.}
        \label{fig:eend_pyannote}
    \end{subfigure}
    \hfill
    \begin{subfigure}[b]{0.49\linewidth}
        \centering
        \includegraphics[width=\textwidth]{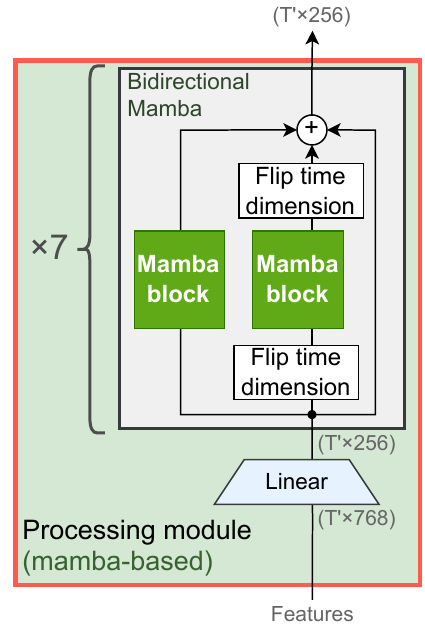}
        \caption{Our mamba-based processing module for the EEND model.}
        \label{fig:eend_bimamba}
    \end{subfigure}
    
    \caption{Local EEND architecture. The activation is respectively sigmoid and softmax for the multilabel and multiclass outputs.}
    \label{fig:eend_architecture}
\end{figure}

The architecture used for the EEND segmentation models is shown in \autoref{fig:eend_pyannote}.
We propose to investigate three main architecture choices: the core processing module, the use of multilabel or multiclass output, and the fixed duration $W$ covered by the local EEND model. 

Inspired by the previous studies~\cite{baroudi2023pyannote, delcroix23_interspeech}, we use a frozen WavLM \cite{wavlm} as our feature extractor. We use the publicly available pretrained ``WavLM Base'' architecture, which extracts 768 features per frame. It ingests audio files at 16kHz of duration $W$ seconds with $T$ samples ($T\mathord{=}16000 \cdot W$). It outputs around 49 frames per second of audio ($T' \approx 49 \cdot W$). The WavLM features are then passed to the processing module. The BiLSTM-based processing module uses the state-of-the-art (SOTA) configuration (4 layers with 128 hidden features) \cite{plaquet23powerset}. The Mamba-based one is detailed in \autoref{sec:mamba-eend}. Finally, the output of the processing module is passed through two linear layers of hidden size 128, and a final linear layer that reduces the number of features to the desired output size $C$. 

We compare the traditional multilabel problem representation against the newer and potentially better multiclass powerset representation~\cite{plaquet23powerset}. With a multilabel output, each label represents the individual activity of one speaker ($C\mathord{=}N$ output labels for $N$ speakers), and training is done using the permutation-free binary cross entropy loss. With a multiclass powerset output, we encode each possible speaker combination as a class ($C\mathord{=}\sum_{i=0}^{K} {N \choose i}$ classes with $N$ speakers and $K$ maximum simultaneous speakers), training is done using the permutation-free cross entropy loss. In both cases, the model can only handle a fixed maximum number of speakers $N$ (and a fixed $K$ for powerset) determined by the architecture. Moreover, we test each of these configurations using local EEND models with window sizes $W\mathord{\in}\{5,10,30,50\}$ seconds.

\subsubsection{Embedding extraction and clustering}

For the second part of the pipeline, we use the process described in \cite{bredin23pyannote21} with a ResNet~\cite{wang2023wespeaker} model pre-trained on VoxCeleb~\cite{VoxCeleb17, VoxCeleb18, VoxCeleb19} for feature extraction. Embeddings are extracted on non-overlapped speech from the output of the local EEND segmentation model. These embeddings are used in an agglomerative hierarchical clustering with centroid linkage, which gives clusters of speaker embeddings (i.e. the global speaker labels). Finally, speaker IDs of all windows are aligned according to the clusters, and overlapped frames of the sliding window are averaged to obtain the global diarization.

\subsection{Proposed Mamba-based segmentation model}
\label{sec:mamba-eend}

Mamba comes from a line of research on State Space Models (SSM) for machine learning. The general principle of these approaches is to learn through a neural network the ``state space representation'' which models a continuous dynamical system. Mamba focuses on an efficient \textit{recurrent} discretization of this continuous system. In practice, Mamba acts as an RNN equipped with selective memorization capabilities (enabling attention-like behavior) and efficient compression of long-term data. 

An individual Mamba block as those used in \autoref{fig:eend_bimamba} possesses a few hyperparameters. We only change the state dimension $d\_state$ (or $N$ in the original paper), which is the size of the internal memory vector, and fix it to $64$ from our preliminary experiments.
For all other parameters, we use the default values of the official \texttt{mamba\_ssm} \footnote{\url{https://github.com/state-spaces/mamba}} python implementation (all identical to the original paper), as we did not find significant improvement from tweaking them in preliminary experiments. 



Our Mamba-based architecture for the segmentation model is shown in \autoref{fig:eend_bimamba} and replaces the BiLSTM of the original pipeline. The number of input features is reduced from 768 to 256 through a Linear layer, this greatly reduces the size of the subsequent BiMamba layers (55.9M $\rightarrow$ 7.4M parameters) which made the model easier to train in our preliminary experiments. The reduced features are then passed to 7 chained BiMamba blocks. We use the \textit{External} Bidirectional Mamba proposed in \cite{zhang2024mambainspeech} which was found to be slightly better than \textit{Internal} BiMamba. The paper also proposes Conformer-like blocks for Mamba, but we were unable to achieve competitive performance with it.

\section{Experiments}

\subsection{Data}


All models are trained on a compound dataset using 8 existing datasets: NOTSOFAR1~\cite{notsofar1}, MSDWild~\cite{msdwild}, VoxConverse~\cite{VoxConverse}, AliMeeting~\cite{AliMeeting}, AMI (channel 1)~\cite{AMI}, MagicData-RAMC~\cite{RAMC}, AISHELL-4~\cite{AISHELL}, and a simulated dataset~\cite{yamashita2022improvingnaturalnesssimulatedconversations} based on LibriSpeech~\cite{librispeech} with MUSAN background noise~\cite{musan} and room impulse response~\cite{ko2017reverberant}.

We do not include DIHARD III \cite{DIHARD} in the compound training set, but we use it for evaluation. We use it for both ``out-of-domain'' evaluation of models trained on the compound dataset, and for evaluating models after domain adaptation to DIHARD III. 

Since the local segmentation model uses a fixed number of speakers $N$, we extract speaker statistics from the training sets to determine what number to use for each window length. For $W\mathord{=}\{5,10,30,50\}$, we choose $N\mathord{=}\{4,4,5,6\}$ respectively and use $K\mathord{=}2$ for powerset models. $N$ and $K$ are selected so that at least 97\% of the data can be correctly processed.


\subsection{Training process}

We train all models for 80 epochs, with 72000 steps using a batch size of 32 (i.e., 1000h for \mbox{$W\mathord{=}50$}, 100h of data for \mbox{$W\mathord{=}5$}). We found this setup to be the fairest to compare different $W$, despite seeing more data, longer windows do not gain an advantage. All datasets are seen equally in training, regardless of their length. We use one warm-up epoch for the learning rate until we reach \mbox{$lr\mathord{=}0.002$}, and then use a cyclic scheduler with a two epochs period, decaying with ten epochs half-life.
For adaptation, we use a fixed \mbox{$lr\mathord{=}0.00005$} and stop the training after 10 epochs with no improvement.

As for the second part of the pipeline, the \textit{clustering threshold} and the \textit{minimum cluster size} are obtained through a hyperparameter search using Optuna~\cite{optuna_2019} for each individual experiment. The search is performed on the compound training and validation sets with Optuna's multivariate Tree-structured Parzen Estimator for 300 iterations.

\subsection{Evaluation of the local segmentation models}
\label{sec:local_der}


The diarization error rate (DER) of the pipeline is heavily influenced by the embedding extraction and clustering step. Specifically, errors made in the local segmentation can either be amplified or corrected, depending on their nature. To fairly compare the quality of the local segmentation of each configuration, we cover each audio file using a sliding window with no overlap, and use \textit{oracle clustering} to stitch them together (for each window, we use the permutation that matches most closely with the ground truth and then aggregate to obtain the final diarization).

\begin{figure}[t]
    \centering
    \includegraphics[width=0.80\linewidth]{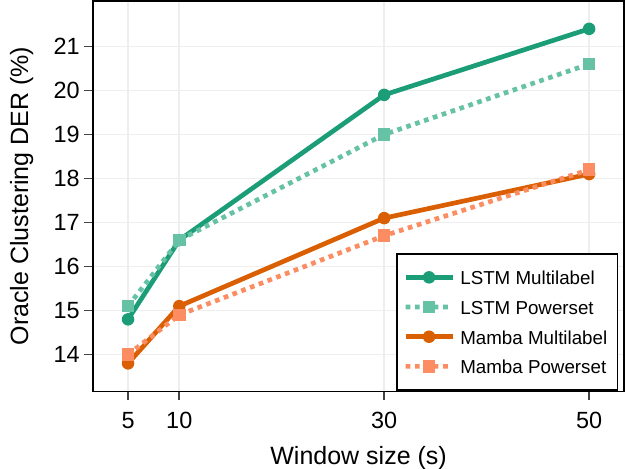}
    \caption{Oracle clustering DER as a function of window size for each architecture. Sliding windows do not overlap.}
    \label{fig:localder}
\end{figure}

\autoref{fig:localder} shows the relationship between window size and DER with oracle clustering for each architecture and loss, clearly showing the impact of all tested factors. First, the oracle clustering DER increases with the window size, this is not surprising as longer windows require the models to keep track of more speakers for longer durations. For all configurations, the speaker confusion is the main DER component that increases with window size.

Second, for any configuration, the Mamba-based models perform much better than the BiLSTM-based ones. The greater difference on longer windows can be attributed to Mamba's superior selective memorization.

Finally, the powerset loss tends to help the models achieve better DER than with the multilabel loss, especially for LSTM-based architectures. It seems that the powerset loss makes the problem \textit{easier to learn}: it helps the most on LSTM (which are relatively weak) with long sequences (harder problem). 

\begin{table}[t]
    \tablefontsize

    \caption{Macro average of the DER of all domains on the validation set (files under 100 seconds are excluded).}
    \label{tab:pipelineder}
    
    \centering
    \begin{tabular}{|ll|rrrr|}
    \cline{3-6}
    \multicolumn{1}{c}{} &  & \multicolumn{4}{c|}{Window size} \\
    \cline{1-2}
    Architecture & Loss type & 5s & 10s & 30s & 50s \\
    \hline
    \hline
    \multirow[t]{2}{*}{LSTM} & Multilabel & 18.1 & \textbf{17.7} & 19.0 & 20.1 \\
     & Powerset & 18.4 & 18.2 & 18.7 & 19.5 \\
    \hline
    \multirow[t]{2}{*}{Mamba} & Multilabel & 17.7 & 16.3 & \textbf{16.1} & 16.9 \\
     & Powerset & 17.6 & 16.6 & 16.4 & 17.0 \\
    \hline
    \end{tabular}
\end{table}

\subsection{Evaluation of the full system on the validation set}
\label{sec:ppl_val}

\begin{table*}[t]
    \caption{DER comparison to September 2024 SOTA. (Parentheses) indicates a 0.25s collar was used for evaluation. \derbg{Bold} indicates the best known DER and \derbl{blue background} indicates the best DER among our systems (we also include systems less than 2\% worse relatively).}
    \label{tab:globalder}

    \tablefontsize
    \setlength\extrarowheight{0.06cm}
    \centering
    \begin{tabular}{|l l|c|c|c|c|c|c|c||>{\columncolor[gray]{0.9}}c||c|}
    \hline
        System & \vthml{Parameters} & \vth{AISHELL-4} & \vthml{AliMeeting\\ \textit{far}} & \vthml{AMI\\ \textit{(channel~1)}} & \vthml{MSDWild \textit{(Few)}} &  \vth{RAMC} & \vthml{NOTSOFAR-1\\\textit{(channel 1)}} & \vthml{VoxConverse\\ \textit{v0.3}} &  \vthml{In-domain\\macro average} & \vth{DIHARD III}\\
    \hline
    (1) LSTM 10s & 96.5M & 11.5 & 21.2 & 20.1 & \margintxt{23.2}{16.6} & 11.5 & 23.8 & \margintxt{9.8}{7.2} & 17.3 & 25.7 \nlt
    (2) \variant Domain adaptation & & 11.0 & 18.7 & 19.1 & \margintxt{21.7}{15.3} & \derbl{\derbg{11.0}} & 25.0 & \margintxt{9.8}{7.3} & 16.7 & 17.9 \nlt
    \hline
    (3) Attention 10s & 102M & 11.9 & 21.7 & 21.3 & \margintxt{24.6}{18.2} & 12.1 & 25.2 &  \margintxt{10.5}{7.8} & 18.2 & 24.9 \nlt
    \hline
    (4) Mamba 10s & 101M & 11.0 & 20.0 & 19.5 & \margintxt{22.5}{16.0} & \derbl{\derbg{11.1}} & \derbl{23.6} & \margintxt{10.1}{7.5} & 16.8 & 24.4 \nlt
    (5) \variant Domain adaptation & & \derbl{\derbg{10.5}} & 17.6 & 19.1 & \margintxt{21.3}{14.8} & \derbl{\derbg{11.1}} & \derbl{23.3} & \margintxt{\derbl{9.3}}{6.7} & 16.0 & \derbl{16.8} \nlt
    (6) Mamba 30s & & 10.8 & 18.0 & 19.0 & \margintxt{20.0}{13.7} & \derbl{\derbg{11.1}} & 25.0 & \margintxt{10.3}{7.7} & 16.3 & 23.5 \nlt
    (7) \variant Domain adaptation & & \derbl{\derbg{10.5}} & \derbl{16.2} & \derbl{18.5} & \margintxt{\derbl{\derbg{19.8}}}{13.6} & 11.4 & 26.9 & \margintxt{9.9}{7.3} & 16.2 & \derbl{16.7} \nlt
    \hline
    SOTA & & \makecell{\derbg{10.6}\\ \cite{baroudi24_interspeech}} & \makecell{\derbg{13.2}\\ \cite{harkonen24_eendm2f}} & \makecell{\derbg{17.1}\\ \cite{kalda2024pixit}} & \parbox{1.5cm}{\centering \margintxt{\derbg{19.6}}{10.0} \\ \cite{baroudi24_interspeech}} & \makecell{\derbg{11.1} \\ \cite{harkonen24_eendm2f}} & ----- & \parbox{1.5cm}{\centering \margintxt{-----}{\derbg{4.0}} \\ \cite{baroudi2023pyannote}} & \diagbox[height=2\line,width=1cm]{}{} & \makecell{\derbg{16.1} \\ \cite{harkonen24_eendm2f}} \\
    \hline
    \end{tabular}
\end{table*}

In order to select the best setting, we evaluate the pipeline DER on the validation set. \autoref{tab:pipelineder} shows the macro average DER of all domains after tuning the clustering hyperparameters. 
Only for these results, we exclude files under 100s from the validation set since longer models might act as (nearly) EEND on short files, this is desirable in real use cases, but it biases the results.

We can observe that all Mamba configurations beat the best LSTM configuration. LSTM obtains better results with $W\mathord{=}10$ and Mamba with $W\mathord{=}30$. In both cases, we find that the powerset representation degrades the DER. Since the influence of window size and powerset loss seem to be the opposite of what was found in \autoref{fig:localder}, we propose to analyze the causes of these differences.

\subsubsection{Impact of the window size}

\autoref{fig:localder} shows oracle clustering DER increases with the window size. However, while longer window sizes result in a worse segmentation than shorter windows, they provide more samples for speaker embedding extraction. This creates a trade-off where a worse segmentation might be offset by a better clustering, which explains why $W\mathord{=}5$ is not strictly better. Although we can identify a ``best $W$'' for both LSTM and Mamba architectures, the former is much more sensitive to $W$ and its DER still degrades quickly with larger values.

\subsubsection{Impact of the powerset formulation}

In \autoref{fig:localder}, the powerset formulation never significantly degrades the oracle clustering DER, but \autoref{tab:pipelineder} shows that it degrades the final DER in most cases.
A lower local DER does not directly translate to a lower pipeline DER because some types of error are more harmful than others. The most impactful are speaker confusion and \textit{overlapped} missed speaker detection, which causes the embeddings to be extracted from audio segments that includes different speakers or overlapped speech. These errors result in noisy embeddings and a poor clustering step. False alarm causes the inclusion of nonspeech (e.g., background noise) or exclusion of speech (if overlapping is detected), which is not as damaging. We observed that local EEND models with powerset loss had a different distribution of errors than the multilabel loss. It has a lower false alarm but a higher speaker confusion, which is detrimental to the speaker embedding extraction and clustering steps and explains the apparent discrepancy.


\subsection{Detailed comparison of the best systems}

We focus on the best LSTM and Mamba configurations previously shown in bold in the \autoref{tab:pipelineder} and compare them on the evaluation sets of seven datasets in \autoref{tab:globalder}. We also compare these systems to one with an attention-based processing module. We report SOTA results on each of these datasets and highlight the best systems.
To compare with SOTA results, we benchmark our systems using a frozen \textit{WavLM Base+} as a feature extractor, before and after domain adaptation (finetuning and pipeline parameter optimization).

The reference LSTM-based system already beats SOTA on AISHELL-4, RAMC, and AliMeeting after domain adaptation (system (2)). Although this architecture had already been used \cite{baroudi2023pyannote}, it was not yet benchmarked on these datasets.

\subsubsection{Comparison to an attention-based architecture}

For system~(3), we use transformer encoder blocks as the processing module instead of LSTMs. We used the code, encoder blocks architecture, learning rate and scheduling from \cite{kinoshita2021icasp} and implemented it into our existing pyannote-based framework. It ends up very similar to our proposed Mamba-based architecture in terms of the number of layers, features, and parameters.

Despite multiple attempts, we found the architecture difficult to train. The transformer-based 10s model was trained for 80 epochs with 1000h per epoch but is still strictly inferior to the LSTM-based model (system~(1)) except on DIHARD III. In comparison, all other architectures using $W=10$ were trained with five times fewer training samples and still achieved much better performance. This does not invalidate the use of attention for this architecture but shows that it might be more complex and costly to train.

\subsubsection{Comparison to the proposed Mamba-based architecture}

To distinguish the impact of the window size from the impact of Mamba, we compare both 10s Mamba-based systems~(4)(5) directly comparable to (1)(2)(3), and a 30s Mamba-based systems~(6)(7), which corresponds to the best configuration found in \autoref{sec:ppl_val}. 

At equal window size $W\mathord{=}10$, the Mamba-based architecture~(4) always equals or surpasses the LSTM~(1). The same goes for their domain-adapted versions (5) and (2). In particular, we observe better performance on the complex DIHARD III dataset. 

 While comparison of 10s mamba (4)(5) and 30s mamba (6)(7) is not as straightforward, we can observe significant improvements using longer windows on the first four in-domain datasets and DIHARD. RAMC, NOTSOFAR-1 and VoxConverse exhibit slightly different trends. Most models obtain similar performance on RAMC, and longer windows do not matter as much since they only contain two-speakers conversations. System~(7) gives a better local EEND DER than (6) but a worse pipeline DER, which means the clustering part is at fault. NOTSOFAR-1 and VoxConverse seems to contain too little data given their complexity (33h and 20h of non evaluation data respectively) to properly finetune models and select the best checkpoint. Moreover, using longer windows effectively reduces the number of training and validation samples, which might not make it the most optimal option for smaller datasets.


Our proposed Mamba-based system achieves SOTA performance on RAMC, AISHELL and MSDWILD, and remains competitive on DIHARD and AMI.

\subsubsection{Impact of the parameter count}

More than 90\% of the parameters in our models are from the frozen WavLM (94M parameters). However, our proposed Mamba-based processing module (8.1M parameters) is larger than the LSTM one (2.1M parameters). To confirm this does not give Mamba an unfair edge, we increased the number of parameters of the LSTM (more layers or hidden features to match those of Mamba) but found it always led to performance degradation. It does not strictly invalidate these LSTM architectures, but shows they quickly become impossibly hard and costly to train.

For example, with a 6.8M parameters BiLSTM (increasing hidden state size $128 \rightarrow 256$) we obtain an in-domain average DER of $18.2\%$ (worse than (1)). We did the reverse with Mamba, decreasing its number of parameters to 2.2M (decreasing features $256 \rightarrow 128$) and found an in-domain average DER of $17.1\%$, which is still better than system (1) despite using a similar number of parameters.


\section{Conclusion}
We investigated the use of Mamba for speaker diarization in an EEND-VC pipeline. We found that Mamba offers a powerful alternative to BiLSTM and attention-based models for speaker diarization. In particular, we found that Mamba-based segmentation models are capable of handling longer window sizes than LSTM, which improves the pipeline performance over shorter windows in most datasets.
The proposed system is simple, but achieves SOTA performance on three datasets and competitive results on DIHARD and AMI.



\clearpage   
\bibliographystyle{IEEEbib-abbrev.bst} 
\bibliography{bibliography} 

\end{document}